\documentclass[graybox]{svmult}

\usepackage{mathptmx}       
\usepackage{helvet}         
\usepackage{courier}        
\usepackage{type1cm}        
\usepackage{makeidx}         
\usepackage{graphicx}        
\usepackage{multicol}        
\usepackage[bottom]{footmisc}

\makeindex             

\newcommand{\cm}{cm$^{-1}$} \newcommand{\A}{\AA$^{-1}$} \newcommand{\Q}{$\vec{Q}$} 

\begin{document}

\title*{The Macroscopic Quantum Behavior of Protons in the KHCO$_3$ Crystal: Theory and Experiments}
\titlerunning{Macroscopic Quantum Entanglement} 
\author{Fran\c{c}ois Fillaux, Alain Cousson and Matthias J. Gutmann}
\institute{Fran\c{c}ois Fillaux \at UPMC Univ Paris 06, UMR 7075, LADIR, F-75005, Paris, France\\ CNRS, UMR7075, LADIR, 2 rue H. Dunant, F-94320, Thiais, France.\\ \email{fillaux@glvt-cnrs.fr}
\and Alain Cousson \at Laboratoire L\'{e}on Brillouin (CEA-CNRS), CE Saclay, 91191 Gif-sur-Yvette, cedex, France. \\ \email{alain-f.cousson@cea.fr}
\and Matthias J. Gutmann \at ISIS Facility, Rutherford Appleton Laboratory, Chilton, Didcot, OX11 0QX, UK. \\ \email{m.j.gutmann@rl.ac.uk}}
%
%
\maketitle

\abstract*{For hydrogen bonded crystals exhibiting proton transfer along hydrogen bonds, namely $\mathrm{O1-H}\cdots \mathrm{O2} \longleftrightarrow \mathrm{O1}\cdots \mathrm{H-O2}$, there is a dichotomy of interpretation consisting in that while the crystal lattice is a quantum object with discrete vibrational states, protons are represented by a statistical distribution of classical particles with definite positions and momenta at any time. We propose an alternative theoretical framework for decoherence-free macroscopic proton states. The translational invariance of the crystal, the adiabatic separation of proton dynamics from that of heavy atoms, the nonlocal nature of proton states, and quantum interferences, are opposed to statistical distributions and semiclassical dynamics. We review neutron scattering studies of the crystal of potassium hydrogen carbonate (KHCO$_3$) supporting the existence of macroscopic quantum correlations, from cryogenic to room temperatures. In addition, quantum fluctuations calculated for superposition states in thermal equilibrium are consistent with measurements of the correlation time. There is no temperature induced transition from the quantum to the classical regime. The crystal can be therefore represented by a state vector and the dichotomy of interpretation must be abandoned. } 

\abstract{For hydrogen bonded crystals exhibiting proton transfer along hydrogen bonds, namely $\mathrm{O1-H}\cdots \mathrm{O2} \longleftrightarrow \mathrm{O1}\cdots \mathrm{H-O2}$, there is a dichotomy of interpretation consisting in that while the crystal lattice is a quantum object with discrete vibrational states, protons are represented by a statistical distribution of classical particles with definite positions and momenta at any time. We propose an alternative theoretical framework for decoherence-free macroscopic proton states. The translational invariance of the crystal, the adiabatic separation of proton dynamics from that of heavy atoms, the nonlocal nature of proton states, and quantum interferences, are opposed to statistical distributions and semiclassical dynamics. We review neutron scattering studies of the crystal of potassium hydrogen carbonate (KHCO$_3$) supporting the existence of macroscopic quantum correlations, from cryogenic to room temperatures. In addition, quantum fluctuations calculated for superposition states in thermal equilibrium are consistent with measurements of the correlation time. There is no temperature induced transition from the quantum to the classical regime. The crystal can be therefore represented by a state vector and the dichotomy of interpretation must be abandoned. }

\section{Introduction}
\label{sec:1}
The linear formalism of quantum mechanics extrapolated from the level of electrons and atoms to that of everyday life leads to conclusions totally alien to commonsense, such as Schr\"{o}dinger's Cat in a superposition of ``alive-dead'' states and nonlocal observables \cite{EPR}. Such conflicts lead to a dichotomy of interpretation consisting in that, while at the microscopic level a quantum superposition indicates a lack of definiteness of outcome, at the macroscopic level a similar superposition can be interpreted as simply a measure of the probability of one outcome or the other, one of which is definitely realized for each measurement of the ensemble \cite{Bohr,Laloe,Leggett1,Leggett,LCDFG,LG}. For open systems, this can be legitimated by decoherence \cite{Zurek} stipulating that an initial superposition state should lose its ability to exhibit quantum interferences via interaction with the environment. However, since the quantum theory does not predict any definite dividing line between quantal and classical regimes, macroscopic quantum behavior is possible for systems decoupled from, or very weakly coupled to, the surroundings \cite{ACAL}. In principle, there is no upper limit in size, complexity, and temperature, beyond which such systems should be doomed to classicality. 

For example, it is a matter of fact that defect-free crystals are macroscopic quantum systems with discrete phonon states at any temperature below melting or decomposition. This is an unavoidable consequence of the translational invariance of the lattice. However, the dichotomy of interpretation arises for the so-called ``proton disorder'' in crystals containing O--H$\cdots$O hydrogen bonds. The coexistence of two configurations at thermal equilibrium, say $\mathrm{O1-H}\cdots \mathrm{O2}$ and $\mathrm{O1}\cdots \mathrm{H-O2}$, has been thoroughly investigated in many systems \cite{SZS}. Although the light mass of protons suggests that dynamics should be quantum in nature, semiclassical approaches are widely used to rationalize correlation times measured with solid-state NMR and quasi-elastic neutron scattering (QENS). Semiclassical protons are thought of as dimensionless particles, with definite positions and momenta, moving in a double-wells coupled to an incoherent thermal bath. These protons undergo uncorrelated jumps over the barrier and ``incoherent tunneling'' through the barrier. In fact, these models describe a liquid-like surroundings, at variance with the spatial periodicity of the crystal, and strong interaction with the thermal bath is supposed to lead to fast decoherence. By contrast, vibrational spectra provide unquestionable evidences that the translational invariance and the quantum nature of lattice dynamics are not destroyed by proton transfer. Our purpose is therefore to elaborate a purely quantum rationale avoiding any mixture of quantum and classical regimes.

We shall concentrate on the crystal of potassium hydrogen carbonate (KHCO$_3$) composed of centrosymmetric dimers of hydrogen bonded carbonate ions (HCO$_3^-)_2$ separated by K$^+$ entities. At elevated temperatures, the coexistence of two configurations for dimers is commonly conceived of as a statistical distribution \cite{BHT,EGS,TTO1,TTO2}. By contrast, systematic neutron scattering experiments measuring a large range of the reciprocal space have revealed the macroscopic quantum behavior of protons, from cryogenic to room temperatures, and the theory suggests that this behavior is intrinsic to the crystal state \cite{Fil3,FCG2,FCKeen,IF}. The present contribution is a preliminary attempt to elaborate a consistent presentation of experimental and theoretical works currently in progress. 

In Sec. \ref{sec:2} we present the crystal structure and the thermally activated interconversion of dimers. We emphasize why this crystal is unique to observing macroscopic quantum effects. In Sec. \ref{sec:3}, we show that the adiabatic separation of proton dynamics leads to decoherence-free states. Then, we introduce the theoretical framework for macroscopic proton states in Sec. \ref{sec:4} and the double-well for protons in Sec. \ref{sec:5}. In Sec. \ref{sec:6}, the calculated scattering cross-section allows us to interpret neutron scattering experiments in terms of quantum correlations. In Sec. \ref{sec:7}, quantum beats arising from the superposition of macroscopic proton states in thermal equilibrium are compared with the correlation time determined with QENS. In the conclusion, we emphasize that the crystal is a macroscopic quantum object that can be represented by a state vector. 

\section{The crystal structure of KHCO$_3$}
\label{sec:2}

\begin{figure}[!hbtp]
\begin{center}
\includegraphics[scale=.2]{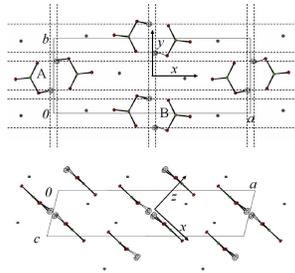}
\end{center}
\caption{\label{fig:01} Schematic view of the crystalline structure of KHCO$_3$ at 14 K. \textit{Dashed lines} through protons are guides for the eyes. $a = 15.06(2)$ \AA, $b = 5.570(15)$ \AA, $c = 3.650(8)$ \AA, $\beta = 103.97(15)^\circ$. The ellipsoids represent 50\% of the probability density for nuclei.}
\end{figure}

The crystal at 14 K is monoclinic, space group P$2_1/a$ (C$_{2h}^5$), with four equivalent KHCO$_3$ entities per unit cell (Fig. \ref{fig:01}). Centrosymmetric dimers (HCO$_3^-)_2$ linked by moderately strong OH$\cdots$O hydrogen bonds, with lengths $R_{\mathrm{O}\cdots\mathrm{O}} \approx  2.58$ \AA, are well separated by K$^+$ ions. All dimers lie practically in (103) planes, hydrogen bonds are virtually parallel to each other, and all protons are crystallographically equivalent (indistinguishable). This crystal is unique to probing proton dynamics along directions $x,\ y,\ z,$ parallel to the stretching ($\nu$OH), the in-plane bending ($\delta$OH), and the out-of-plane bending ($\gamma$OH) vibrational modes, respectively. 

\begin{figure}[!hbtp]
\begin{center}
\includegraphics[scale=0.6]{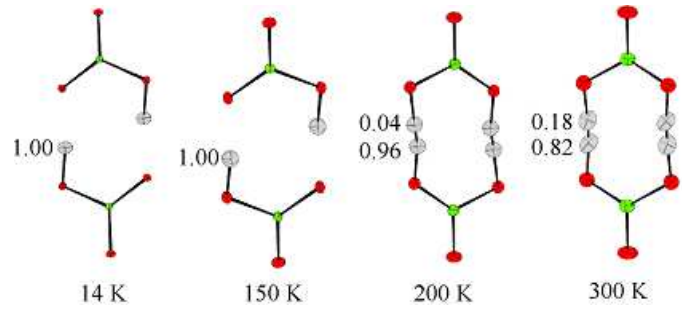}
\end{center}
\caption{\label{fig:02} Probability densities for protons in the KHCO$_3$ crystal at various temperatures, determined with single-crystal neutron diffraction. The ellipsoids represent 50\% of the probability density for nuclei.}
\end{figure}

From 14 K to 300 K, there is no structural phase transition. The increase of the unit cell dimensions and of the hydrogen bond length are marginal, but the population of proton sites changes significantly (Fig. \ref{fig:02}). Below $\approx 150$ K, all dimers are in a unique configuration, say $L$. At elevated temperatures, protons are progressively transferred along the hydrogen bonds to the less favored sites (configuration $R$) at $\approx 0.6$ \AA\ from the main position. The center of symmetry is preserved and all proton sites remain indistinguishable. There is a general agreement that the population of the less favored site (or interconversion degree $\varrho$) is determined by an asymmetric double-well potential along the hydrogen bonds \cite{BHT,EGS,Fil2,FTP}, but an in-depth examination of proton dynamics is necessary to distinguishing statistical disorder or quantum delocalization. 

\section{\label{sec:3}The adiabatic separation}

Within the framework of the Born-Oppenheimer approximation, the vibrational Hamiltonian can be partitioned as
\begin{equation}\label{eq:1}
\mathcal{H}_\mathrm{v} = \mathcal{H}_{\mathrm{H}} +\mathcal{H}_{\mathrm{at}}+ \mathcal{C}_{\mathrm{Hat}},
\end{equation}
where $\mathcal{H}_{\mathrm{H}}$ and $\mathcal{H}_{\mathrm{at}}$ represent the sublattices of protons (H$^+$) and heavy atoms, respectively, while $\mathcal{C}_{\mathrm{Hat}}$ couples the subsystems. For OH$\cdots$O hydrogen bonds, coupling terms between OH and O$\cdots$O degrees of freedom are rather large \cite{Novak,SZS}, and beyond the framework of the perturbation theory. Two approaches, either semiclassical or quantum, are commonly envisaged.

In the semiclassical view, protons are thought of as dimensionless particles, with definite positions and momenta, moving across a potential hypersurface \cite{BVIT,BVT,GPNK,SFS2,TVL}. Complex trajectories involving heavy atoms lead to mass renormalization, and to incoherent phonon-assisted tunnelling \cite{BHT,EGS,ST}. This approach is quite natural when the Born-Oppenheimer surface is calculated from first principles, but quantum effects can be severely underestimated. 

Alternatively, if the classical concept of ``trajectory'', totally alien to quantum mechanics, is abandoned, adiabatic separation of the two subsystems, namely $\mathcal{H}_{\mathrm{H}}$ and $\mathcal{H}_{\mathrm{at}}$, may lead to tractable models \cite{FCG2,FRLL,GPNK,MW,SZS,W}. Then, light protons in a definite eigen state should remain in the same state in the course of time, while heavy atoms oscillate slowly, in an adiabatic hyperpotential depending on the proton state, through the coupling term. This separation is relevant for KHCO$_3$ because adiabatic potentials for different protons states do not intersect each other. Then, protons are bare fermions and quantum correlations should occur \cite{FCG2}. 

In fact, the separation is rigorously exact in the ground state, since protons should remain in this state for ever, if there is no external perturbation. Furthermore, for asymmetric double-wells, with wave functions largely localized in each well (see below Sec. \ref{sec:5}), the adiabatic separation should also hold for the lowest state of the upper minimum and long-lived superposition states should interfere. 

\section{\label{sec:4}Macroscopic proton states}

Consider a crystal composed of very large numbers $N_\mathrm{a}$, $N_\mathrm{b}$, $N_\mathrm{c}$ ($\mathcal{N}=N_\mathrm{a}N_\mathrm{b}N_\mathrm{c}$) of unit cells labelled $j,k,l,$ along crystal axes $(a),$ $(b),$ $(c)$, respectively. The two dimers per unit cell are indexed as $j,k,l$ and $j',k,l$, respectively, with $j = j'$. For centrosymmetric dimers, there is no permanent dipolar interaction, so that interdimer coupling terms and phonon dispersion are negligible \cite{FTP,IKSYBF,KIN}. The eigen states of the sublattice of protons can be therefore represented in a rather simple way with the basis sets of eigen states for isolated dimers 

A H1--H2 dimer is modelled with coupled centrosymmetric collinear oscillators in three dimensions, along coordinates $\alpha_{1\mathrm{jkl}}$ and $\alpha_{2\mathrm{jkl}}$ ($\alpha =x,y,z$), with respect to the center at $\alpha_{0\mathrm{jkl}}$. The mass-conserving normal coordinates independent of $j, k,l,$ and their conjugated momenta, 
\begin{equation}\label{eq:2}
  \begin{array}{lc}
\alpha_{\mathrm{s}} = \displaystyle{\frac {1} {\sqrt{2}}\left(\alpha_{1} - \alpha_{2} + 2\alpha_0 \right)}, & P_{\mathrm{s}\ualpha} = \displaystyle{\frac {1} {\sqrt{2}} \left( P_{1\ualpha} - P_{2\ualpha} \right)}, \\
    \alpha_{\mathrm{a}} = \displaystyle{\frac {1} {\sqrt{2}}\left(\alpha_{1} + \alpha_{2} \right)}, & P_{\mathrm{a}\ualpha} = \displaystyle{\frac {1} {\sqrt{2}}\left( P_{1\ualpha} + P_{2\ualpha } \right)},
  \end{array}
\end{equation}
lead to uncoupled oscillators at frequencies $\hbar\omega_{\mathrm{s}\ualpha}$ and $\hbar\omega_{\mathrm{a}\ualpha}$, respectively, each with $m = 1$ amu. The difference $(\hbar\omega_{\mathrm{s}\ualpha} - \hbar\omega_{\mathrm{a}\ualpha})$ depends on the coupling term (say $\lambda_\ualpha$). The wave functions, $\Psi_{\mathrm{njkl}}^\mathrm{a}( \alpha_{\mathrm{a}})$, $\Psi_{\mathrm{n'jkl}}^\mathrm{s}(\alpha_{\mathrm{s}} -\sqrt{2}\alpha_{0})$, cannot be factored into wave functions for individual particles, so there is no local information available for these entangled oscillators. Consequently, the degenerate ground state of indistinguishable fermions must be antisymmetrized. For this purpose, the wave function is rewritten as a linear combination of those for permuted oscillators as
\begin{equation}\label{eq:3}
\Theta_{0\mathrm{jkl}\pm } = \displaystyle{\frac{1} {\sqrt{2}}} \prod\limits_\ualpha \Psi_{0\mathrm{jkl}}^\mathrm{a} (\alpha_{\mathrm{a}}) \left [ \Psi_{0\mathrm{jkl}}^\mathrm{s}(\alpha_{\mathrm{s}} -\sqrt{2}\alpha_{0}) \pm \Psi_{0\mathrm{jkl}}^\mathrm{s} (\alpha_{\mathrm{s}} +\sqrt{2}\alpha_{0}) \right ],
\end{equation}
and the antisymmetrized state vectors as: 
\begin{equation}\label{eq:4}
\begin{array}{rcl}
|0jkl+ \rangle \otimes |S\rangle & = & \big | \Theta_{0\mathrm{jkl}+} \rangle \otimes \displaystyle{\frac{1}{\sqrt{2}}} \left [ |\uparrow_1 \downarrow_2 \rangle - | \downarrow_1 \uparrow_2 \rangle \right ] ;\\
|0jkl- \rangle \otimes |T\rangle & = & | \Theta_{0\mathrm{jkl}-} \rangle \otimes \displaystyle{\frac{1}{\sqrt{3}}} \left [ | \uparrow_1 \uparrow_2 \rangle + | \downarrow_1 \downarrow_2 \rangle + \displaystyle{\frac{1}{\sqrt{2}}} [ |\uparrow_1 \downarrow_2 \rangle + |\downarrow_1 \uparrow_2 \rangle ] \right ] .
\end{array} 
\end{equation}
The oscillators are now entangled in position, momentum, and spin. In contrast to magnetic systems \cite{Cowley}, there is no level splitting, so the symmetry-related entanglement is energy-free. It is also independent of $\lambda_\alpha$. Furthermore, contrariwise to Keen and Lovesey \cite{KL1}, or Sugimoto et al. \cite{SOY}, we argue, as an experimental fact, that there is no significant exchange integral for protons separated by $\approx 2.2$ \AA\ \cite{FCou}. Neutron diffraction and spectroscopy show that protons in KHCO$_3$ are neither delocalized nor itinerant particles and there is no sizable energy band structure. 

In quantum mechanics, normal coordinates (\ref{eq:2}) define nonlocal pseudoprotons ($m = 1$ amu), say $\mathcal{P}_{\mathrm{sjkl}}$ and $\mathcal{P}_{\mathrm{ajkl}}$, with an internal degree of freedom corresponding to symmetric or antisymmetric displacements of two ``half-protons'', respectively. Each dimer site is a superposition of two such half-protons. Obviously, pseudoprotons are totally alien to the intuitive conception of particles, based on classical mechanics, but they are the actual observables, whereas individual particles are not. 

Consider now the sublattice of protons. The spatial periodicity leads to collective dynamics and nonlocal observables in three dimensions. With the vibrational wave function for the unit cell $j,\ k,\ l,$ namely $\Xi_{0\mathrm{jkl}\utau } = \Theta_{0\mathrm{jkl}\utau } \pm \Theta_{0\mathrm{j'kl}\utau }$, where $\tau =$ ``$+$'' or ``$-$'', usual phonon waves can be written as 
\begin{equation}\label{eq:5}
\Xi_{0\utau} (\vec{k})= \displaystyle{\frac{1}{\sqrt {\mathcal{N}}}}  \sum\limits_{\mathrm{l} = 1}^{\mathrm{N_c}} \sum\limits_{\mathrm{k} = 1}^{\mathrm{N_b}} \sum\limits_{\mathrm{j} = 1}^{\mathrm{N_a}} \Xi_{0\mathrm{jkl}\utau} \exp(\I\vec{k\cdot L}), 
\end{equation}
where $\vec{k}$ is the wave vector and $\vec{L}  = j \vec{a} + k \vec{b} + l \vec{c}$, with the unit cell vectors $\vec{a}$, $\vec{b}$, $\vec{c}$. This equation represents collective dynamics of H--H dimers thought of as composed bosons. This would be correct if the crystal was composed of indistinguishable dimer entities (KHCO$_3)_2$. However, neutron diffraction shows that the crystal structure is composed of KHCO$_3$ entities related to each other through the appropriate symmetry operations. The probability density of each atom is equally distributed over all equivalent sites and, conversely, the probability density at each site includes contributions from all indistinguishable atoms of the same kind. Consequently, the sublattice of protons must be thought of as a sublattice of nonlocal indistinguishable fermions and antisymmetrization of the plane waves (\ref{eq:5}) leads to 
\begin{equation}\label{eq:6}
\vec{k\cdot L} \equiv 0 \mathrm{\ modulo\ } 2\pi. 
\end{equation}
Consequently, there is no phonon (no elastic distortion) in the ground state and this symmetry-related ``super-rigidity'' \cite{FCG2} is totaly independent of proton--proton interaction. Then, the lattice state vectors in three dimensions can be written as: 
\begin{equation}\label{eq:7}
\begin{array}{c}
\left | \Xi_{0 +} (\vec{k = 0}) \right \rangle \otimes |S \rangle; \\
\left | \Xi_{0 -} (\vec{k = 0}) \right \rangle \otimes |T \rangle. 
\end{array}
\end{equation}
Each macroscopic state of the sublattice represents a nonlocal pseudoparticle with a mass $m = 1$ amu, namely a pseudoproton, $\mathcal{P}_a$ or $\mathcal{P}_s$, with a definite spin-symmetry and an occupation number of $(4\mathcal{N})^{-1}$ per site. There is no local information available for these entangled states and the wave functions $\Xi_{0 \utau} (\vec{k = 0})$ represent collective oscillations of the super-rigid lattice as a whole with respect to the center of mass of the crystal. Finally, the ground state of the sublattice is a superposition of the pseudoproton states as:
\begin{equation}\label{eq:8}
\begin{array}{c}
\sqrt{\mathcal{N}} | \Xi_{0 +} (\vec{k = 0}) \rangle \otimes |S \rangle;\\ 
\sqrt{\mathcal{N}} | \Xi_{0 -} (\vec{k = 0}) \rangle \otimes |T \rangle. 
\end{array}
\end{equation}
This ground state is intrinsically steady against decoherence. Irradiation by plane waves (photons or neutrons) may single out some excited pseudoprotons. Entanglement in position and momentum is preserved, while the spin-symmetry and super-rigidity are destroyed. However, the spin-symmetry is reset automatically after decay to the ground state, presumably on the time-scale of proton dynamics. Consequently, disentanglement reaches a steady regime such that the amount of transitory disentangled states is determined by the ratio of the density-of-states for the surrounding atmosphere and external radiations, on the one hand, and for the crystal, on the other. This ratio is so small that disentangled states are too few to be observed, but they allow the super-rigid sublattice to be at thermal equilibrium with the surroundings, despite the lack of internal dynamics. The main source of disentanglement is actually the thermal population of excited proton states. However, even at room temperature, the thermal population of the first excited state ($< 1\%$ for $\gamma$OH $\approx 1000$ \cm) is of little impact to effective measurements. 

For the sublattice of bosons in the isomorph crystal of KDCO$_3$, (\ref{eq:3}) and (\ref{eq:6}) are not relevant. There is neither spin-symmetry nor super-rigidity. Dynamics are represented with normal coordinates (\ref{eq:2}) and phonons (\ref{eq:5}). Needless to say, the H and D atoms have the same number of 12 degrees of freedom per unit cell, but the symmetrization postulate shrinks the size of the allowed Hilbert space from $\sim 12^\mathcal{N}$ for bosons to $\sim 12\mathcal{N}$ for fermions. 

\section{\label{sec:5}Proton dynamics}

\begin{figure}[!hbtp]
\begin{center}
\includegraphics[scale=0.4]{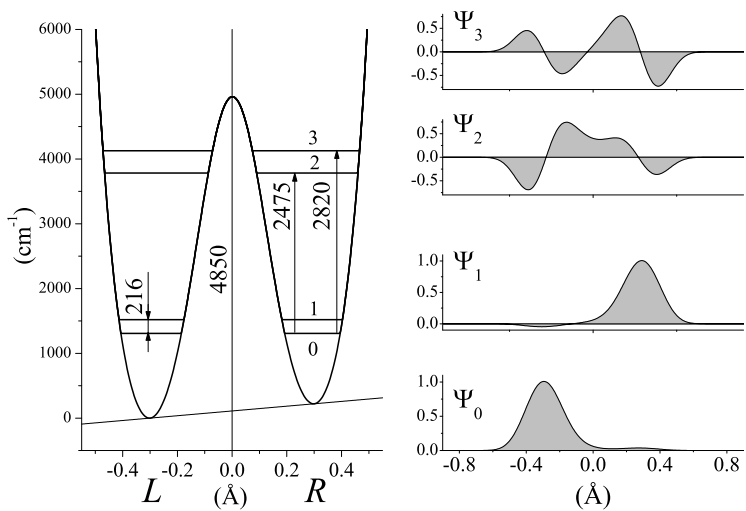}
\end{center}
\caption{\label{fig:03} Potential function and wave functions for the OH stretching mode along the hydrogen bond in the KHCO$_3$ crystal. $V(x) = 374x + 0.4389\times 10^6 + 5516 \exp(-30.8x^2)$. $V$ and $x$ are in \cm\ and \AA\ units, respectively \cite{Fil2,FTP}. The oscillator mass is 1 amu. }
\end{figure}

The interconversion degree at thermal equilibrium (Fig. \ref{fig:02}) is determined by the potential function for protons. On the one hand, the bending modes do not play any significant role, since they show rather modest anharmonicity and the population of excited states is negligible. On the other hand, the double-wells for the OH stretching (Fig. \ref{fig:03}) is known from experiments. The distance between minima ($2x_0 \approx 0.6$ \AA) is given by the crystal structure. The upper states at $h \nu_{02}$ and $h \nu_{03}$ were determined from infrared and Raman band profiles \cite{Fil2,Fil1}, and the ground state splitting ($h \nu_{01}$) was observed with incoherent inelastic neutron scattering (IINS) \cite{FTP,KIN}. The potential obtained through best fitting exercises is over determined and largely model independent. In addition, the oscillator mass of 1 amu is not a free parameter. It is determined by $2x_0$ for a given set of energy levels. 

For the $|0\rangle$ and $|1\rangle$ states, the potential asymmetry leads to substantial localization of the wave functions in the lower and upper wells, respectively. However, tunneling is possible through the tiny delocalized fraction ($\varepsilon \approx 0.05$) visible in Fig. \ref{fig:03}. 

\begin{figure}[!hbtp]
\begin{center}
\includegraphics[scale=0.6]{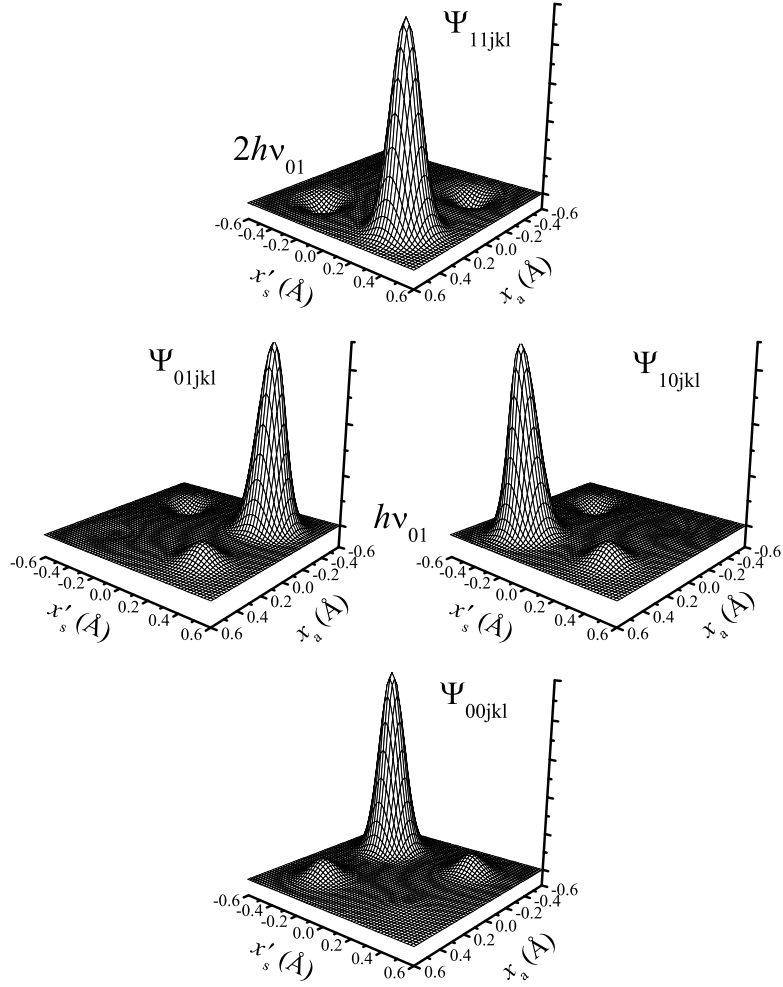}
\end{center}
\caption{\label{fig:04} Schematic view of the tunneling wave functions. For the sake of clarity, the weak component of the wave function in one dimension is multiplied by a factor of 2. }
\end{figure}

This potential has been a puzzle ever since it was determined because the upper minimum was naively thought of as corresponding to the transfer of a proton. However, this is unlikely, for this would lead to unrealistic dimers composed of di-protonated (H$_2$CO$_3$) and non-protonated (CO$_3^{2-}$) entities \cite{FLR}. Such entities are ruled out by the centrosymmetric character of proton dynamics established by the symmetry-related selection-rules observed in the infrared and Raman \cite{LN}. It is now clear that, if pseudoprotons are \textit{the} observables, this nonlocal potential accounts for pseudoproton dynamics along $x_\mathrm{a}$ or $x_\mathrm{s}$. The $|0\rangle \longleftrightarrow |1\rangle$ transition corresponds to the through-barrier transfer (tunneling) of a pseudoproton as a rigid entity, with no energy transfer to the internal degree of freedom. The IINS bandwidth, very close to the spectrometer resolution \cite{FTP}, shows that this transition is virtually dispersion-free and there is no visible splitting suggesting any difference for the transfer of $\mathcal{P}_\mathrm{a}$ or $\mathcal{P}_\mathrm{s}$ \cite{FTP,KIN}. On the other hand, the upper states $|2\rangle$ and $|3\rangle$ correspond to excitations of internal stretching coordinates, $\nu_\mathrm{a}$ (infrared) or $\nu_\mathrm{s}$ (Raman). They are slightly different, but this is unimportant for interconversion since thermal populations are strictly negligible for these states. 

The interconversion dynamics involving $\mathcal{P}_a$ and $\mathcal{P}_s$ can be rationalized with the potential surface along coordinates $x_\mathrm{a}$ and $x'_{\mathrm{s}} = x_{\mathrm{s}} \pm \sqrt{2} x_{0}$: 
\begin{equation}\label{eq:9}
\mathcal{V}(x_{\mathrm{a}},x'_{\mathrm{s}}) = V(x_{\mathrm{a}}) + V(x'_{\mathrm{s}}). 
\end{equation}
\begin{figure}[!hbtp]
\begin{center}
\includegraphics[scale=0.6]{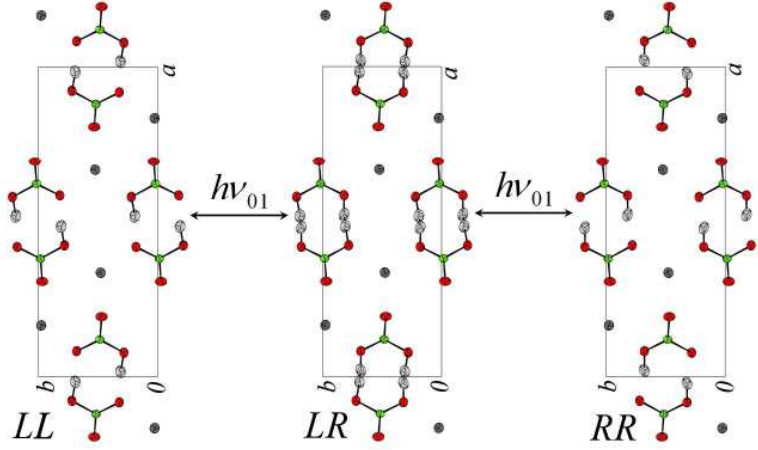}
\end{center}
\caption{\label{fig:05} Schematic view of proton configurations for the tunnelling states. In the ground state $LL$ and in the upper state $RR$ at $2h\nu_{01}$, protons are fully entangled. In the intermediate state $LR$ (or $RL$) at $h\nu_{01}$ all proton sites are equally occupied. The spin symmetry and the super-rigidity are destroyed. }
\end{figure}
The energy level scheme comprises three states at 0, $h\nu_{01}$ (with twofold degeneracy) and $2h\nu_{01}$. The non-symmetrized local wave functions (Fig. \ref{fig:04}) are simple product of the local wave functions in one dimension (Fig. \ref{fig:03}). Proton configurations for the three states are tentatively sketched in Fig. \ref{fig:05}. The ground state corresponds to the structure observed at low temperature, with both $\mathcal{P}_\mathrm{a}$ and $\mathcal{P}_\mathrm{s}$ in the $L$ configuration. The antisymmetrized macroscopic state analogous to (\ref{eq:7}), namely $|0+\rangle|S\rangle |0-\rangle |T\rangle$, can be obtained via (\ref{eq:3}) to (\ref{eq:6}). Similarly, the upper state vector at $2h \nu_{01}$ with both pseudoprotons in the $R$ configuration is $|1+\rangle|S\rangle |1-\rangle|T\rangle$. In the intermediate state at $h \nu_{01}$, only one pseudoproton (either $\mathcal{P}_\mathrm{a}$ or $\mathcal{P}_\mathrm{s}$) is transferred to the $R$ configuration, so the spin-symmetry and the super-rigidity are destroyed. Then, plane waves (\ref{eq:5}) lead to state vectors $|1+,0-,\vec{k}_{10}\rangle$ and $|0+,1-, \vec{k}_{01}\rangle$. Note that the $|0\rangle \longleftrightarrow |1\rangle$ transition is effectively observed with IINS, thanks to energy and momentum transfer, whereas the $|0\rangle \longleftrightarrow |2\rangle$ transition cannot be probed directly, according to the quantum theory of measurements. 

\begin{figure}[!hbtp]
\sidecaption[t]
\includegraphics[scale=0.3]{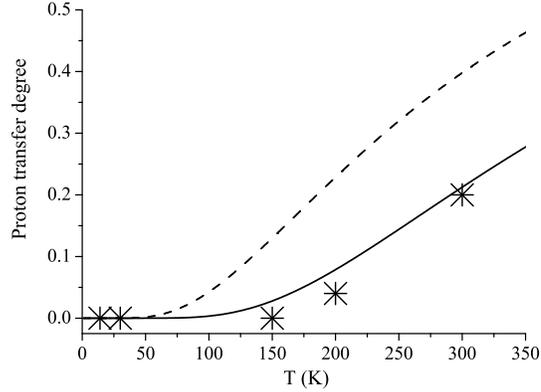}
\caption[t]{\label{fig:06} Temperature effect on the proton transfer degree in the KHCO$_3$ crystal. \textit{Stars}: experimental according to \cite{FCG2}. \textit{Solid line}: eq. (\ref{eq:12}) for two-levels. \textit{Dashed line}: eq. (\ref{eq:10}) for three-levels.}
\end{figure}

The proton transfer degree calculated supposing the three levels at thermal equilibrium is 
\begin{equation}\label{eq:10}
\varrho(T) = [p_{01} (T) + 2p_{01}^2 (T)] [1 + p_{01} (T) + p_{01} (T)^2]^{-1}, 
\end{equation}
where $p_{01} (T) = \exp(-h\nu_{01}/kT)$ is the probability for the transfer of a pseudoproton. The dashed line in Fig. \ref{fig:06} clearly shows that this equation is at variance with observations.

In fact, owing to the adiabatic separation of proton dynamics, energy exchange with the surroundings occurs exclusively via photons, with the momentum conservation rule $\vec{k}_{10} + \vec{k}_{01} \equiv \vec{0}$. Consequently, re-entanglement occurs spontaneously in the intermediate state as 
\begin{equation}\label{eq:11}
\begin{array}{l}
|1+,0-,\vec{k}_{10}\rangle + |0+,1-,\vec{k}_{01}\rangle = 2^{-1/2} [|0+\rangle |S\rangle |0-\rangle|T\rangle + |1+\rangle |S\rangle |1-\rangle |T\rangle ]
\end{array}
\end{equation}
and the transfer degree 
\begin{equation}\label{eq:12}
\varrho(T) = 2p_{01}^2 (T) [1 + p_{01}^2 (T)]^{-1},
\end{equation}
is in reasonably good agreement with measurements (see the solid line in Fig. \ref{fig:06}). The energy difference of $2h\nu_{01}$ between $RR$ and $LL$ configurations is therefore confirmed and Figure \ref{fig:06} is an indirect evidence that the degenerate intermediate state observed with IINS is not thermodynamically stable, thanks to the purely quantum re-entanglement mechanism (\ref{eq:11}). 

However, the interpretation of Bragg diffraction is ambiguous because the $LL$ and $RR$ configurations of the proton sublattice are crystallographically equivalent, as they are related through a translation vector $(a/2,b/2,0)$ (see Fig. \ref{fig:05}). The reciprocal lattices are identical and it is unknown whether neutrons were diffracted by either sublattice, with probability $1-\varrho$ and $\varrho$, respectively, or by a superposition state, $(1- \varrho)^{1/2} |0+\rangle |S\rangle |0-\rangle |T\rangle + \varrho^{1/2} |1+\rangle |S\rangle |1-\rangle |T\rangle$. The former case is a mixture of $LL$ and $RR$ configurations analogous to disorder under consideration in many crystallographic \cite{KY,TTO1,TTO2}, solid-state NMR \cite{BHT,Odin} and QENS \cite{EGS} works. Alternatively, a superposition should give rise to quantum interferences corresponding to coherent fluctuations of the probability density at proton sites. 

\section{\label{sec:6}Probing quantum entanglement with neutrons}

Neutrons (spin $1/2$) are unique to observing the spin-symmetry of macroscopic states (\ref{eq:7}). However, quantum entanglement is extremely fragile, because it is not stabilized by any energy. Consequently, only ``noninvasive'' experiments, free of measurement-induced decoherence, are appropriate \cite{LG}. For neutron scattering, this means (i) no energy transfer (ii) no spin-flip and (iii) particular values of the neutron momentum transfer vector \Q\ preserving the super-rigidity. (By definition, $\vec{Q} = \vec{k}_i - \vec{k}_f $, where  $\vec{k}_i$ and $\vec{k}_f$ are the initial and final wave vectors, respectively.) 

The dotted lines in Fig. \ref{fig:01} enhance the network of double-lines of proton sites in dimer planes. We present below neutron scattering experiments for (i) double-lines of protons, (ii) arrays of double-lines in two dimensions, and (iii) the sublattice in three dimensions. For the sake of simplicity, it should be born in mind that elastic scattering events are identical for $LL$ and $RR$ configurations and, therefore, independent of the interconversion degree. 

\subsection{\label{sec:61}Double-lines of entangled protons}

\begin{figure}[!htbp]
\begin{center}
\includegraphics[scale=0.23]{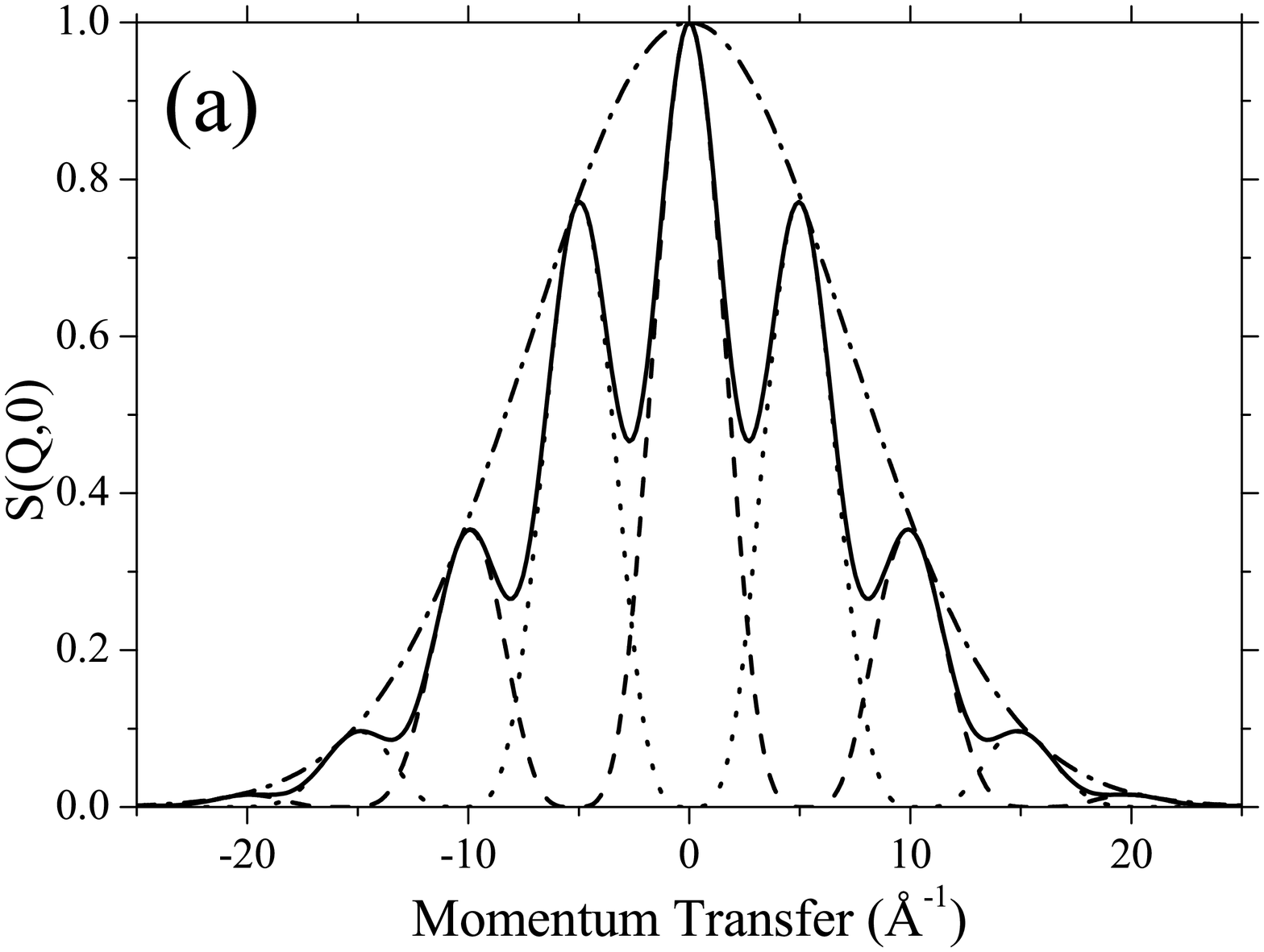}
\includegraphics[scale=0.23]{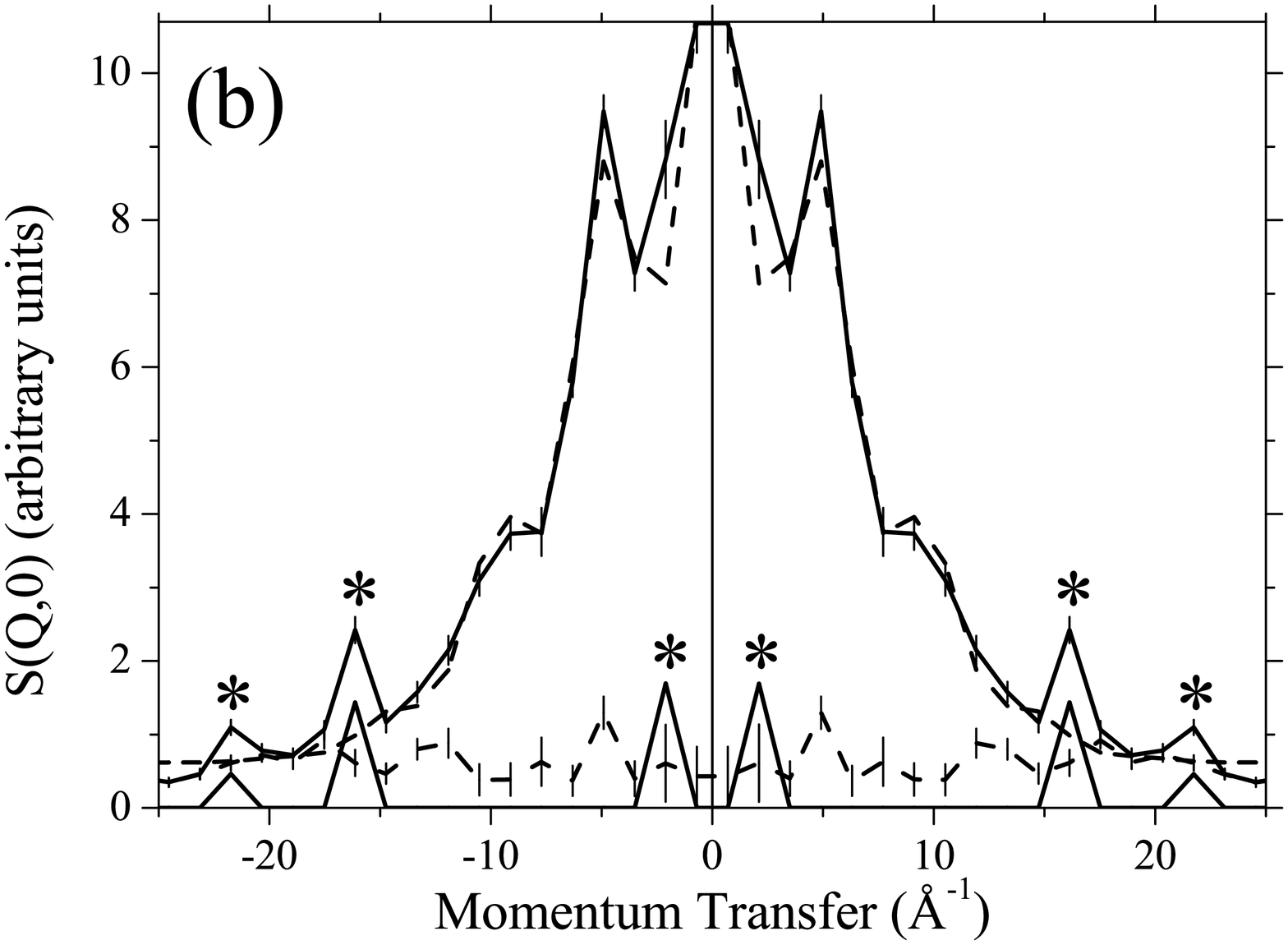}
\end{center}
\caption{\label{fig:07} (a): Comparison of the theoretical profiles $S(Q ,0)$ for a non-entangled pair (\textit{dot-dashed}), and for an entangled pair of fermions, according to (\ref{eq:11}) (\textit{solid line}). The \textit{doted} and \textit{dashed} curves represent interference fringes for in-phase and out-of-phase scattering. (b): $S(Q_y,0)$ measured at 20 K for a single-crystal of KHCO$_3$ (\textit{solid curve with error bars}). Comparison with the best fit (\textit{dashed line}) obtained with (\ref{eq:14}) convoluted with a triangular resolution function. The \textit{dashed line with error bars} is the difference spectrum. *: Triangular functions due to other scattering events.}
\end{figure}

Consider an incoherent elastic neutron scattering (IENS) experiment conducted with (i) the best resolution in energy, in order to reject inelastic scattering events, (ii) a modest resolution in \Q, so Bragg peaks merge into a continuum and long-range correlations are overlooked. For momentum transfer $Q_\ualpha$ along $\alpha$, the scattering function for an entangled pair (\ref{eq:4}) can be written as: 
\begin{equation}\label{eq:13}
\begin{array}{l}
S_{\utau_\mathrm{i}\utau_\mathrm{f}} \left( {Q_\ualpha}, \omega_\ualpha \right) = \sum\limits_{\utau_\mathrm{f}} \sum\limits_{\utau_\mathrm{i}} \\
| \langle 0jkl\tau_\mathrm{i} | \exp \I Q_\ualpha \left( \alpha_{2\mathrm{jkl}} - \alpha_{0\mathrm{jkl}} \right) +  \tau_\mathrm{i} \tau_\mathrm{f} \exp \I Q_\ualpha \left( \alpha_{2\mathrm{jkl}} + \alpha_{0\mathrm{jkl}} \right) | 0jkl\tau_\mathrm{f} \rangle \\
 \times \langle 0jkl\tau_\mathrm{f} | \exp \I Q_\ualpha \left( \alpha_{1\mathrm{jkl}} - \alpha{_{0\mathrm{jkl}}} \right ) + \tau_\mathrm{f} \tau_\mathrm{i} \exp \I Q_\ualpha  \left( \alpha_{1\mathrm{jkl}} + \alpha{_{0\mathrm{jkl}}} \right) | 0jkl\tau_\mathrm{i} \rangle  |^2 \\
\times  \exp(-2W_{\mathrm{L}\ualpha}) \delta(\omega_\ualpha). \\
\end{array}
\end{equation}
Each bracket represents a scattering event by a pseudoproton located at both sites ($\pm \alpha_{0\mathrm{jkl}}$). The product of two brackets means that each neutron is scattered simultaneously by the two pseudoprotons superposed at the same sites, either in-phase, $|\pm\rangle \longleftrightarrow |\pm \rangle$ ($\tau_\mathrm{f} \tau_\mathrm{i} = +1$), or anti-phase, $|\pm\rangle \longleftrightarrow |\mp\rangle$ $(\tau_\mathrm{f} \tau_\mathrm{i} = -1)$. The spin-symmetry is probed along the neutron-spin direction with 100\% probability and there is no spin-flip because the initial and final states are $|0jkl\tau_{\mathrm{i}} \rangle$ and $|0jkl\tau_{\mathrm{f}} \rangle$, respectively, for one scattering event and vice-versa for the other one. Thanks to adiabatic separation, the lattice Debye-Waller factor $\exp(-2W_{\mathrm{L}\ualpha})$ can be factored. The energy transfer is $\hbar\omega_\ualpha$ and $\delta(\omega_\ualpha)$ accounts for energy conservation. In the harmonic approximation, the scattering function is  \cite{Fil3,IF}
\begin{equation}\label{eq:14}
\begin{array}{l}
 S_{\pm\pm} \left(Q_\ualpha, \omega_\ualpha \right) = \\
\cos ^4 \left(Q_\ualpha \alpha_0 \right ) \left[ \exp \displaystyle{- \frac{Q_\ualpha^2 u_{0\ualpha}^2} {\sqrt{1 + 4\lambda _\ualpha }}} + \exp -Q_\ualpha^2 u_{0\ualpha}^2 \right] \exp(-2W_{\mathrm{L}\ualpha}) \delta (\omega_\ualpha),\\
S_{\pm\mp} \left(Q_\ualpha, \omega_\ualpha \right) = \\
\sin^4 \left(Q_\ualpha \alpha_0 \right ) \exp \displaystyle{ \left[ - Q_\ualpha^2 \left( \frac{u_{0\ualpha}^2} {2\sqrt{1 + 4\lambda _\ualpha }} + \frac{{u_{0\ualpha}^2 }} {2} \right) \right] } \exp(-2W_{\mathrm{L}\ualpha}) \delta (\omega_\ualpha).
\end{array} 
\end{equation}
Here, $u_{0\ualpha}^2  = {\hbar }/ {(2m\omega _{0\ualpha} )}$ is the mean square amplitude for uncoupled harmonic oscillators in the ground state. The intensity is proportional to the incoherent nuclear cross-section for protons, $\sigma_\mathrm{Hi} \approx 80.26$ b (1 barn $= 10^{-24}$ cm$^2$), and the gaussian-like profiles for uncorrelated scatterers are modulated by $\cos^4 (Q_\alpha \alpha_0)$ and $\sin^4(Q_\alpha \alpha_0)$ (Fig. \ref{fig:07} a). 

Such interference fringes were effectively observed with the MARI spectrometer \cite{MARI} at the ISIS pulsed-neutron source (Fig. \ref{fig:07} b) \cite{IF}. Best fit exercises yield $\alpha_0$-values in reasonable accordance with the crystal structure and the estimated oscillator mass is virtually equal to 1 amu. These experiments are positive evidences of pseudoproton states with spin-symmetry (\ref{eq:4}). 

In fact, neutron plane waves are scattered coherently by double-lines of entangled pairs perpendicular to $Q_\ualpha$, which are reminiscent of Young's double-slits. However, the interference fringes (\ref{eq:14}) are clearly different from those anticipated for distinguishable (classical) double-slits, which should be proportional to $\cos^2 (Q_\ualpha \alpha_0)$ \cite{DNR,ZGSTM,Zeil}. Equation (\ref{eq:14}) is also at variance with the scattering function for protons delocalized in a symmetric double-wells. In this case, there is no definite spin-symmetry for the tunneling states $|0+\rangle$ (ground state) and $|0-\rangle$ (at $\hbar\omega_\mathrm{t}$). The scattering functions for elastic scattering $|0\pm\rangle \longleftrightarrow |0\pm\rangle$ and inelastic scattering $|0\pm\rangle \longleftrightarrow |0\mp\rangle$ are 
\begin{equation}\label{eq:15}
\begin{array}{rcl}
 S_{\pm\pm} \left(Q_\ualpha, \omega_\ualpha \right) & = & \cos ^2 \left(Q_\ualpha \alpha_0 \right ) \exp \left( - Q_\ualpha^2 u_{0\ualpha}^2 -2W_{\mathrm{L}\ualpha} \right) \delta (\omega_\ualpha),\\
S_{\pm\mp} \left(Q_\ualpha, \omega_\ualpha \right) & = & \sin ^2 \left(Q_\ualpha \alpha_0 \right ) \exp \left( - Q_\ualpha^2 u_{0\ualpha}^2 -2W_{\mathrm{L}\ualpha} \right) \delta (\omega_\ualpha - \omega_\mathrm{t}).\\
\end{array} 
\end{equation}
Then, interferences evidence that a single proton is located in two wells. They are visible if the instrument can effectively resolve the tunnel splitting. Otherwise, complementary fringes would merge into the gaussian profile anticipated for a single-well. 

Clearly, scattering by a superposition of entangled double-lines with spin correlations cannot be confused with other double-slits experiments. The fringes are evidences of nonlocal pseudoprotons and there is no means whatever to probe the local particle behavior. 

Since (\ref{eq:13}) and (\ref{eq:14}) hold in the same way for $LL$ and $RR$ configurations, interferences are independent of the interconversion degree. However, at elevated temperatures, the intensity at large $Q_\ualpha$-values is depressed by the lattice Debye-Waller factor and fringes are less visible. 

Needless to say, interferences are neither expected, nor observed, for KDCO$_3$ \cite{IF}.

\subsection{\label{sec:62} Diffraction}

A necessary condition for noninvasive neutron diffraction is that $Q_\mathrm{x},\ Q_\mathrm{y},\ Q_\mathrm{z}$, should match a node of the reciprocal sublattice of protons, so neutrons probe super-rigid states without any induced distortion. The only information conveyed by such events is the perfect periodicity of the sublattice, so the Debye-Waller factor is equal to unity at any temperature. In addition, thanks to the spin-symmetry, the scattered intensity is proportional to the total cross-section $\sigma_\mathrm{H} \approx 82.0$ b \cite{FCG2,SWL}. Otherwise, the spin-symmetry is destroyed, so the intensity scattered by protons is proportional to the coherent cross-section $\sigma_{\mathrm{Hc}} \approx 1.76$ b and to the Debye-Waller factor for non-rigid lattices. The enhancement factor $\sigma_{\mathrm{H}} / \sigma_{\mathrm{Hc}} \approx 45$ is quite favorable to observing quantum correlations. Furthermore, the intensity scattered by the sublattice of heavy atoms, proportional to $\sigma_{\mathrm{cKCO_3}} \approx 27.7$ b, is depressed by the Debye-Waller factor $\exp-2W_{\mathrm{KCO_3}}(\vec{Q})$. Therefore, the contribution of heavy atoms at large \Q-values is rather weak, compared to that of the entangled sublattice, especially at elevated temperatures. 

The dashed lines in Fig. \ref{fig:01} show that proton sites are aligned along $x$ and $y$, but not along $z$. Consequently, the noninvasive condition can be realized for $Q_\mathrm{x}$ and $\ Q_\mathrm{y}$ exclusively, whereas $Q_\mathrm{z}$ never coincides with a nod of the reciprocal lattice of protons. Then, the diffraction pattern depends on whether we consider incoherent or coherent scattering along $Q_\mathrm{z}$. 

\subsubsection{\label{sec:621} Super-rigid arrays in two dimensions}

Consider diffraction by super-rigid arrays in (103) planes and incoherent scattering along $Q_\mathrm{z}$. In the unit cell, there are two indistinguishable double-lines parallel to $y$, so the periodicity of the grating-like structure is $D_\mathrm{x}/2$, with $D_\mathrm{x} \approx a/\cos 42 ^\circ \approx 20.39$ \AA. On the other hand, the spatial periodicity of double-lines parallel to $x$ is $D_\mathrm{y} = b$. The differential cross-section for a superposition of pseudoproton states (\ref{eq:7}) can be then written as 
\begin{equation}\label{eq:16}
\begin{array}{l}
\displaystyle{\frac{\D\sigma_{2}}{\D\Omega}} \propto \sum\limits_{\mathrm{l} =1} ^{\mathrm{N_c}} \sum\limits_{\utau_{i}} \sum\limits_{\utau_{\mathrm{f}}} \left | \sum\limits_{\mathrm{j} = 1} ^{\mathrm{N'_a}} \sum\limits_{\mathrm{k} =1} ^{\mathrm{N_b}} \left\{ \left [ \exp \I Q_\mathrm{y} \left(kD_\mathrm{y} - y_{0} \right) \right. \right. + \left. \utau_{\mathrm{i}} \utau_{\mathrm{f}} \exp \I Q_\mathrm{y} \left(kD_\mathrm{y} + y_{0} \right ) \right ] \right. \\
\times \left. \left [ \exp \I Q_\mathrm{x} \left(jD_\mathrm{x}/2 - x_{0} \right ) + \utau_{\mathrm{i}}\utau_{\mathrm{f}} \exp \I Q_\mathrm{x} \left(jD_\mathrm{x}/2 + x_{0} \right ) \right ] \right\}^2 \bigg |^2 \exp-2W_\mathrm{z} (Q_\mathrm{z}), 
\end{array} 
\end{equation}
with $N'_\mathrm{a} = 2N_\mathrm{a}$. Neutrons are scattered either in-phase ($\tau_\mathrm{f} \tau_\mathrm{i} = +1$) or anti-phase $(\tau_\mathrm{f} \tau_\mathrm{i} = -1)$ by orthogonal pairs of lines separated by $2x_0 \approx 0.6$ \AA\ and $2y_0 \approx 2.209$ \AA, respectively. The phase matching condition, namely $x_0$ ($y_0$) commensurable with $D_\mathrm{x}/2$ ($D_\mathrm{y}$), is intrinsic to the crystal structure. $\{\cdots\}^2$ accounts for simultaneous scattering by the superposed pseudoproton states, without neutron-spin flip. The compound Debye-Waller factor $\exp-2W_\mathrm{z} (Q_\mathrm{z})$, including contributions from all atoms, corresponds to incoherent scattering along $Q_\mathrm{z}$. The permutation $x_0 \longleftrightarrow -x_0, y_0 \longleftrightarrow -y_0$ gives the same equation for $RR$ and $LL$ configurations. 

\begin{figure}[t]
\includegraphics[angle=0.,scale=0.65]{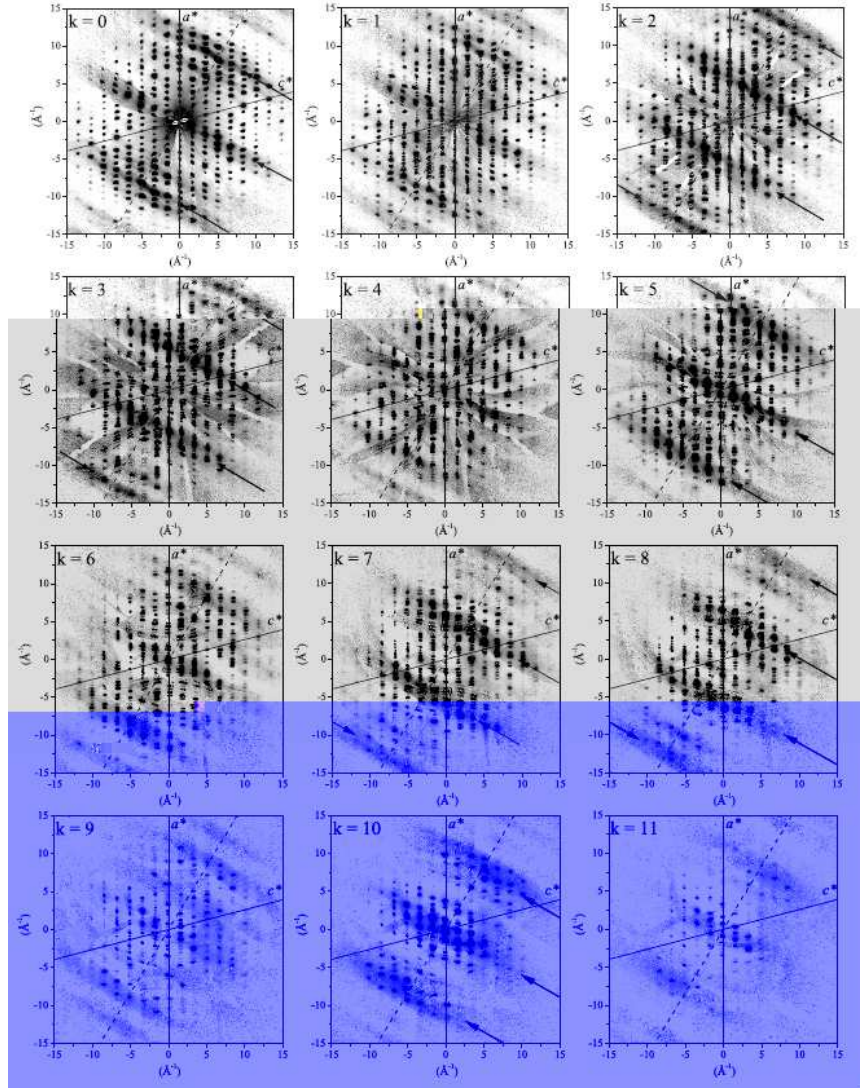} 
\caption{\label{fig:08} Cuts of the diffraction pattern of KHCO$_3$ at 300 K in various ($a^*, c^*$) planes. The arrows emphasize ridges of intensity parallel to $Q_\mathrm{z}$ and perpendicular to the dimer plane (dash lines along $Q_\mathrm{x}$).}
\end{figure}

\begin{table}[!hbtp]
\caption{\label{tab:1} Orders $n_\mathrm{y}$, $n_\mathrm{x}$, and positions $Q_\mathrm{y}$, $Q_\mathrm{x}$, of rods of intensity arising from the entangled array of orthogonal doubles lines of protons in two dimensions. Obs. $Q_\mathrm{x}$: positions in \A\ of the observed rods of intensity along $Q_\mathrm{z}$ in Figs \ref{fig:08} and \ref{fig:09}. $Q_\mathrm{y} D_\mathrm{y}/\pi$ is rounded to integers.}
\begin{tabular}{p{1cm}p{1.5cm}p{1.5cm}p{1cm}p{1.5cm}p{2.5cm}l}
\hline\noalign{\smallskip}
 $n_\mathrm{y}$ & $Q_\mathrm{y}$ (\A)& $Q_\mathrm{y}D_\mathrm{y}/\pi$ & $\tau_\mathrm{i} \tau_\mathrm{f}$ & $k = Q_\mathrm{y}/b^*$ & $Q_\mathrm{x}$ & Obs. $Q_\mathrm{x}$ (\A) \\
\noalign{\smallskip}\svhline\noalign{\smallskip}
0 & 0    & 0    & $+1$ & 0 & $n_\mathrm{x}\pi/x_0$ & $0,\pm 10$\\
1 & 2.86 & 5    & $-1$ & 2.57 & $(n_\mathrm{x}+1/2)\pi/x_0$ & $\pm 5, \pm 15$\\
2 & 5.71 & 10   & $+1$ & 5.14 & $n_\mathrm{x}\pi/x_0$ & $0,\pm 10$\\
3 & 8.57 & 15   & $-1$ & 7.71 & $(n_\mathrm{x}+1/2)\pi/x_0$ & $\pm 5, \pm 15$\\
\noalign{\smallskip}\hline\noalign{\smallskip}
\end{tabular}
\end{table}

The diffraction pattern is composed of rods of diffuse scattering parallel to $Q_\mathrm{z}$, cigar-like shaped by the Debye-Waller factor, at $Q_\mathrm{x},\ Q_\mathrm{y}$-values corresponding to divergences of (\ref{eq:16}). Such divergences occur at $Q_\mathrm{y} = n_\mathrm{y} \pi /y_0 \approx n_\mathrm{y} \times 2.86$ \A\ (Table \ref{tab:1}), since $Q_\mathrm{y} D_\mathrm{y}/\pi \approx 5 n_\mathrm{y}$ is integer. Contrariwise, there is no divergence for anti-phase scattering at $Q_\mathrm{y} = \pm (n_\mathrm{y} + 1/2) \pi/y_0$, because $Q_\mathrm{y} D_\mathrm{y}/\pi \approx 5 (n_\mathrm{y} + 1/2)$ is not integer. For $n_\mathrm{y}$ even, $Q_\mathrm{y} D_\mathrm{y}/\pi$ is also even, $\tau_{\mathrm{i}} = \tau_{\mathrm{f}}$, and ridges are anticipated at $Q_\mathrm{x} = n_\mathrm{x} \pi/x_0 \approx n_\mathrm{x}\times 10$ \A, since $Q_\mathrm{x} D_\mathrm{x}/\pi \approx 68n_\mathrm{x}$ is even. Alternatively, for $n_\mathrm{y}$ odd, $Q_\mathrm{y}D_\mathrm{y} / \pi$ is also odd, $\tau_{\mathrm{i}} \neq \tau_{\mathrm{f}}$, and ridges are anticipated at $Q_\mathrm{x} = (n_\mathrm{x} + 1/2) \pi/x_0$, since $Q_\mathrm{x} D_\mathrm{x}/\pi \approx 68(n_\mathrm{x} + 1/2) $ is even. 

\begin{figure}[!htbp]
\includegraphics[angle=0.,scale=0.33]{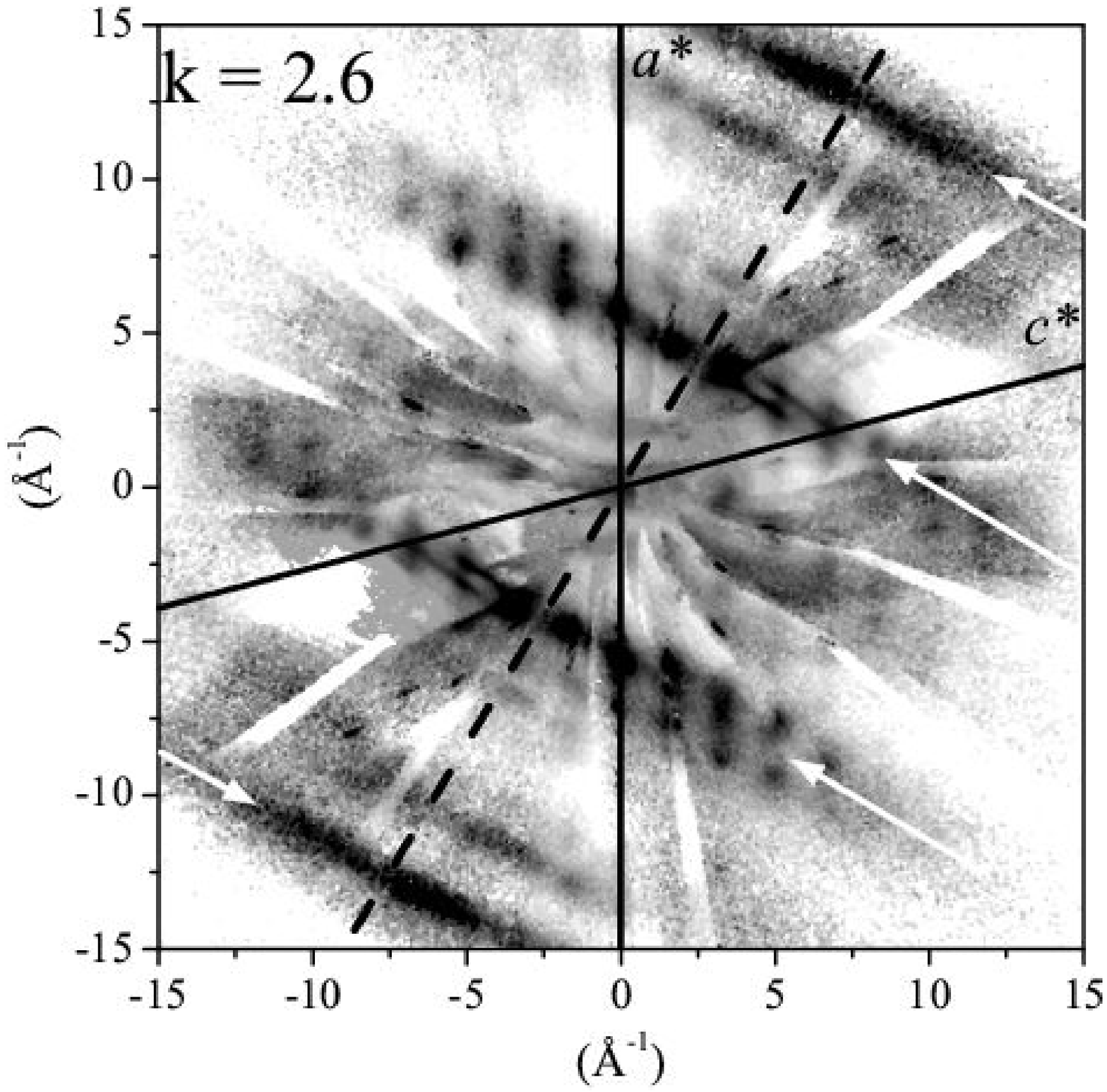} \includegraphics[angle=0.,scale=0.33]{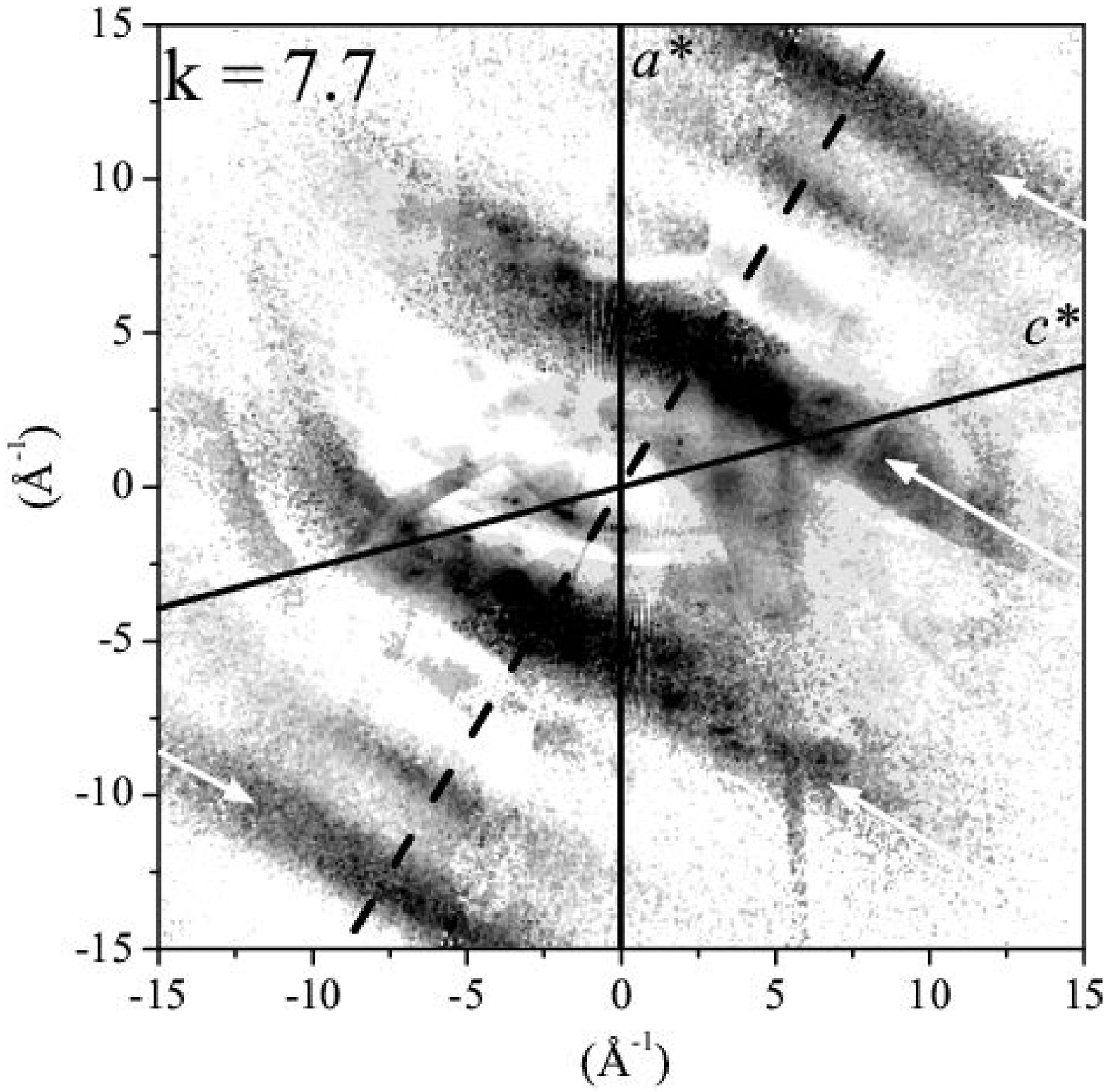}
\caption{\label{fig:09} Diffuse scattering of KHCO$_3$ at 300 K in between  ($a^*, c^*$) reciprocal planes. The arrows emphasize ridges of intensity parallel to $Q_\mathrm{z}$ and perpendicular to the dimer plane (dashed lines along $Q_\mathrm{x}$).}
\end{figure}

\begin{figure}[!htbp]
\sidecaption[t]
\includegraphics[angle=0.,scale=0.35]{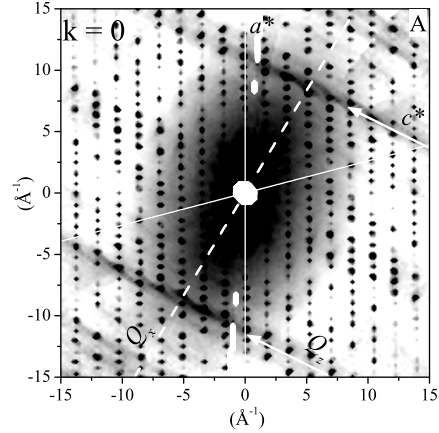}
\caption{\label{fig:10} Diffraction pattern of KHCO$_3$ at 30 K in the ($a^*, c^*$) reciprocal plane at $k = 0$. The arrows emphasize ridges of intensity parallel to $Q_\mathrm{z}$ and perpendicular to the dimer plane (dash lines along $Q_\mathrm{x}$).}
\end{figure}

The cigar-like shaped rods were effectively observed with the SXD \cite{SXD,SXD2} instrument at the ISIS pulsed neutron source (Figs \ref{fig:08}--\ref{fig:10}). For $k = 0$, they appear at $Q_\mathrm{x} = 0$ and $\pm (10.00 \pm 0.25)$ \A, in accordance with $2x_0 \approx 0.6$ \AA. For $k = 1$, they are barely visible. For $k = 2$ or 3, we observe ridges at $Q_\mathrm{x} = \pm (5 \pm 0.2)$ and $\pm (15 \pm 0.2)$ \A, still along $Q_\mathrm{z}$. These features are best visible in Fig. \ref{fig:09} for $k = 2.6$, in accordance with Table \ref{tab:1}. There is no visible ridge at $k = 4$. Then, from $k = 5$ to 9, we observe in Fig. \ref{fig:08} the same sequence as for $k = 0$ to 4, and rods at $k=7.7$ in Fig. \ref{fig:09}. 

Similar diffuse scattering was observed at low temperature in the $k = 0$ plane, at 14 K \cite{FCKeen} and 30 K \cite{FCG2} (Fig. \ref{fig:10}). As anticipated, the rods are unaffected by the interconversion degree. However, at low temperatures, they are partially hidden by the anisotropic diffuse intensity, centered at \Q\ $= \vec{0}$, due to elastic and inelastic incoherent scattering by protons. This continuum precludes observation of the ridges at $\pm 5$ \A\ for $2 \le k \le 3$. Quite paradoxically, quantum correlations are best visible at elevated temperatures. 

As anticipated from Sec. \ref{sec:4}, the same experiments performed with a crystal of KDCO$_3$ do not evidence any cigar-like shaped ridge of enhanced intensity, in addition to regular Bragg peaks, for the sublattice of bosons \cite{FCKeen}. 

\subsubsection{\label{sec:622} Super-rigid arrays in three dimensions}

From (\ref{eq:16}), the differential cross-section for the $LL$ or $RR$ pseudoproton states is written as 
\begin{equation}\label{eq:17}
\begin{array}{rcl}
\displaystyle{\frac{\D\sigma_{3}}{\D\Omega}} & \propto & \sum\limits_{\utau_{\mathrm{i}}} \sum\limits_{\utau_{\mathrm{f}}} \left | \sum\limits_{\mathrm{j} = 1} ^{\mathrm{N'_a}} \sum\limits_{\mathrm{k} =1} ^{\mathrm{N_b}} \sum\limits_{\mathrm{l} =1} ^{\mathrm{N_c}} \right. \left\{ \left [ \exp \imag Q_\mathrm{y} \left(kD_\mathrm{y} - y_{0} \right)\right. + \left. \utau_{\mathrm{i}}\utau_{\mathrm{f}} \exp \I Q_\mathrm{y} \left(kD_\mathrm{y} + y_{0} \right ) \right ] \right. \\
& \times & \left. \left [\exp \I Q_\mathrm{x} \left(jD_\mathrm{x}/2 - x_{0} \right ) + \utau_{\mathrm{i}}\utau_{\mathrm{f}} \exp \I Q_\mathrm{x} \left(jD_\mathrm{x}/2 + x_{0} \right ) \right ] \exp \I Q_\mathrm{z} l D_\mathrm{z} \right\}^2 \bigg |^2 \\
\end{array} 
\end{equation}
This equation describes no spin-flip scattering events that do not destroy the spin-symmetry. Divergences occur along the previous rods of intensity at $Q_\mathrm{z} = \pm n_\mathrm{z} 2\pi / D_\mathrm{z}$, with $D_\mathrm{z} \approx c\times \cos 28^\circ \approx 3.28$ \AA\ and $2 \pi /D_\mathrm{z} \approx 1.92$ \A. These enhanced peaks are visible in Figs \ref{fig:08}--\ref{fig:10}, even at rather large $Q_\mathrm{z}$-values, thanks to super-rigidity. At elevated temperatures, they clearly emerge from the rods of diffuse scattering (\ref{eq:16}) depressed by the Debye-Waller factor. These enhanced peaks were not observed for the deuterated crystal \cite{FCKeen}.

Experiments presented above, in this present section, are clearly consistent with pseudoprotons forming decoherence-free macroscopic single-particle states with spin-symmetry. The underlying theoretical framework presented in Sec. \ref{sec:4} is based on fundamental laws of quantum mechanics. There is no ad hoc hypothesis or parameter. The adiabatic separation, clearly validated by observations, can be regarded as an intrinsic property of hydrogen bonds in this crystal. 

\section{\label{sec:7}Quantum interferences}

Superposition of decoherence-free proton states must lead to quantum interferences, or quantum beats. The non-antisymmetrized wave functions for the states $|0\rangle$ and $|1\rangle$ (Fig. \ref{fig:03}) can be written as \cite{FLR}: 
\begin{equation}\label{eq:18}
\begin{array}{rlll}
\Psi_{0\mathrm{jkl}} & = & \cos\phi\ \psi_0(x-x_\mathrm{m}) & + \sin\phi\ \psi_0(x + x_\mathrm{m});\\
\Psi_{1\mathrm{jkl}} & = - & \sin\phi\ \psi_0(x-x_\mathrm{m}) & + \cos\phi\ \psi_0(x + x_\mathrm{m});\\
\end{array}
\end{equation}
where $x$ stands for $x_\mathrm{a}$ or $x'_\mathrm{s}$ and $\psi_0(x \pm x_\mathrm{m})$ are harmonic eigen functions for the second-order expansion of the potential around the minima at $\pm x_\mathrm{m}$; $\tan 2\phi = \nu_{0\mathrm{t}}/(\nu_{01} - \nu_{0\mathrm{t}})$, where $h\nu_{0\mathrm{t}} \approx 18$ \cm\ is the tunnel splitting for the symmetric potential. Then, $\cos\phi \approx 1$ and $\sin\phi = \varepsilon \approx 5\times 10^{-2}$. Superposition leads to harmonic oscillations of the probability density at the beating frequency $\nu_{\mathrm{0b}}  = 8\varepsilon\nu_{01} \approx 4\nu_{0\mathrm{t}} \approx 2.5\times 10^{12}$, in proton per second units ($\mathrm{H\ s}^{-1}$) \cite{CTDL,Fil7,FLR}.

The $LL \longleftrightarrow RR$ fluctuation rate can be rationalized with two distinct mechanisms, either single-step or two-stepwise (Fig. \ref{fig:04}). The single-step mechanism corresponds to superposition of the states at 0 and $2h\nu_{01}$, corresponding to $LL$ and $RR$ configurations (Fig. \ref{fig:05}), respectively. The interconversion rate due to quantum beats is:
\begin{equation}\label{eq:19}
\nu_{1\mathrm{b}} = 2\varepsilon\nu_{0\mathrm{b}} \exp(-2h\nu_{01}/kT). 
\end{equation}
For the two-stepwise process, firstly, either $\mathcal{P}_\mathrm{a}$ or $\mathcal{P}_\mathrm{s}$ is transferred at $\vec{k = 0}$ to the $LR$ configuration (Fig. \ref{fig:05}) with probability $\exp(-h\nu_{01}/kT)$. Secondly, this state undergoes fast re-entanglement (\ref{eq:11}) leading to the upper state (configuration $RR$) with probability $\exp(-2h\nu_{01}/kT)$. The interconversion rate is then  
\begin{equation}\label{eq:20}
\nu_{2\mathrm{b}} = 2\nu_{0\mathrm{b}} \exp(-3h\nu_{01}/kT). 
\end{equation}

These theoretical rates must be compared to QENS measurements of a KHCO$_3$ crystal, from 200 to 400 K \cite{EGS}. The scattering geometry (\Q\ $\parallel x$) was selected in order to probe proton dynamics specifically along the hydrogen bonds. The inverse relaxation time (or attempt frequency), $\tau_0^{-1} = 2\times 10^{12} \mathrm{s}^{-1}$, is sufficiently close to $\nu_{0\mathrm{b}}$ to suggest that (i) QENS and vibrational spectroscopy techniques probe the same dynamics and, (ii) the two-stepwise mechanism (\ref{eq:20}) is prevailing, in accordance with the larger pre-factor. In addition, the measured rate follows an Arrhenius law with an activation energy $E_\mathrm{a} = (336\pm32)$ \cm\ significantly different from $3h\nu_{01} \approx 648$ \cm. In fact, (\ref{eq:20}) accounts for coherent fluctuations of two pseudoprotons, with probability $\exp(-3h\nu_{01}/kT)$, and pre-factor $2\nu_{0\mathrm{b}}$, while QENS probes the fluctuation rate of a pseudoproton with probability $\exp(-3h\nu_{01}/2kT)$, and pre-factor $\nu_{0\mathrm{b}}$. Hence, $3h\nu_{01}/2 = 324$ \cm\ accords with $E_\mathrm{a}$. It transpires that the QENS technique is an incoherent probe of coherent oscillations of the proton probability, because neutrons are plane waves, rather than a particle-like probe of incoherent stochastic jumps \cite{EGS}. 

In fact, semiclassical models \cite{BHT,EGS,ST} are based on inappropriate premises. (i) The potential asymmetry supposedly due to static effects of neighbouring dimers should be temperature dependent. This is at variance with the interconversion degree (\ref{eq:12}) and Fig. \ref{fig:06}. (ii) Coupling to phonons is posited to be necessary to mediate the through-barrier proton transfer at low temperatures. This is not relevant within the framework of the adiabatic separation. (iii) A smooth transition to the Arrhenius behaviour of classical jumps is supposed to occur at elevated temperatures. Contrariwise, neutron diffraction shows that there is no transition to the classical regime (Figs \ref{fig:08}--\ref{fig:10}). In addition, (\ref{eq:19}) and (\ref{eq:20}) show that an Arrhenius behavior is not necessarily an evidence of the semiclassical regime. Logically, these incorrect premises lead to confusing the activation energy $E_a$ with the potential barrier \cite{EGS}.

\section{Conclusion}

It is often argued that a complex system in continuous interaction with its environment should be in a significantly mixed state that cannot be represented by a state vector. In marked contrast to this widespread opinion, we have accumulated consistent experimental evidences that the sublattice of protons can be represented by a state vector at any temperature up to 300 K. This macroscopic object exhibits all features of quantum mechanics: nonlocality, entanglement, superposition and quantum interferences. There is no transition to the classical regime because the plane waves of the thermal bath cannot destroy entanglement intrinsic to the lattice periodicity. The spin-symmetry of proton states can be transitorily destroyed but the decoherence degree is insignificant because the density-of-states of the surroundings is negligible compared to that of the crystal. 

The cornerstones of the theoretical framework are: (i) adiabatic separation; (ii) the fermionic nature of protons; (iii) indistinguishability and degeneracy. There is no ad hoc hypothesis or parameter. Entanglement is intrinsic to the crystal symmetry, irrespective of the strength of proton--proton interactions. Dynamics is rationalized with pseudoprotons forming macroscopic single-particle states with remarkable spin-symmetry and super-rigidity. These quantum correlations are effectively probed with neutrons and quantum interferences arising from entangled double-lines or long-range correlations in three dimensions emphasize that protons in the crystal field are not individual particles possessing properties on their own right. Collective dynamics suggest that the whole crystal should be conceived of as a matter field that is a superposition of macroscopic single-pseudoparticle states. In addition, super-rigidity adds a new item, along with superfluidity and superconductivity, to the list of quantum ``super'' properties in the condensed matter. 

The interconversion degree at thermal equilibrium is consistent with the double-well potentials for pseudoprotons determined from vibrational spectra and quantum beats accord with QENS measurements of the fluctuation rate. The double-wells for protons is invariant over the whole range of timescales from $\nu$OH vibrations ($\sim 10^{-15}$ s) to diffraction, through QENS, and at any temperature. 

This review emphasizes that the dichotomy of semiclassical protons in a quantum crystal lattice should be abandoned. There is every reasons to suppose that this conclusion holds for many hydrogen bonded crystals and macroscopic quantum behaviors open up new vistas for further investigations.

\end{document}